\documentclass{article}
\usepackage{spconfa4,amsmath,graphicx}
\usepackage{booktabs}  
\usepackage{float}     
\usepackage{array}   
\usepackage{placeins}
\usepackage{subcaption}
\usepackage{multirow} 
\usepackage{cite}
\usepackage{hyperref}
\usepackage{balance}
\usepackage{etoolbox}
\apptocmd{\thebibliography}{\interlinepenalty=10000\relax}{}{}

\title{Mask-Based Speech Enhancement for Spatial Audio: A Comparison of Ambisonics, Beamforming, and Microphone Channels}

\twoauthors
  {Sheli Hendel and Boaz Rafaely}
	{Ben-Gurion University of the Negev\\
	School of Electrical and Computer Engineering\\
	Beer-Sheva 84105, Israel}
  {Dorothea Kolossa}
	{Technische Universität Berlin\\
    Electronic Systems of Medical Engineering\\
	Berlin 10623, Germany}

\begin{document}
%
\maketitle
\thispagestyle{empty}
\begingroup\renewcommand{\thefootnote}{}
\footnotetext{Funded by Deutsche Forschungsgemeinschaft (DFG, German Research Foundation) – Project number KO3434/9-1.}

\begin{abstract}
Mask-based speech enhancement is widely used for suppressing noise and interference, but its performance in spatial audio algorithms with multichannel output has not been studied extensively. In such settings, speech enhancement must improve speech quality while preserving spatial cues that are essential for localization, spatial awareness, and spatial release from masking. In this work, we systematically compare time–frequency masking applied to three signal representations: microphone signals, beamformer outputs, and Ambisonics signals. Performance is evaluated in terms of speech quality, intelligibility, binaural cue preservation, and reverberation preservation. Results reveal a clear trade-off between enhancement and spatial fidelity: beamformer-domain masking achieves the highest speech enhancement scores, while Ambisonics-domain masking better preserves the spatial attributes of the residual interference. All methods preserve the target's localization cues.
\end{abstract}
\section{Introduction}
\label{sec:intro}

Speech enhancement improves speech quality and intelligibility in noisy environments. Masking-based methods, ranging from classical time–frequency filtering to deep learning approaches, have become a dominant paradigm~\cite{lugasi2020speech,rafaely2022spatial,tokala2022binaural,wang2025lightweight,sun2019deep}. While many studies employ multichannel inputs, the enhanced output is typically single-channel. Furthermore, the objective is mainly to suppress interference while maximizing speech quality or intelligibility, often measured using metrics such as Perceptual Evaluation of Speech Quality (PESQ), Short-Time Objective Intelligibility (STOI), or scale-invariant signal-to-distortion ratio (SI-SDR)~\cite{tokala2022binaural,huang2014deep,huang2015joint}.

Recently, masking-based speech enhancement has been extended to multichannel spatial audio applications, including immersive communication, augmented and virtual reality (AR and VR), hearing assistance, and spatial reproduction~\cite{herzog2023ambisep,wang2025lightweight,sun2019deep,leng2021compromise,kumar2025relative}. In these applications, the processing must not only suppress interferers but also preserve spatial cues that are essential for sound localization, scene awareness, listener comfort, and spatial release from masking (SRM)~\cite{lugasi2020speech,blauert2013technology,hauth2020modeling,rafaely2022spatial}. However, masking may distort inter-channel spatial information, degrading localization and spatial perception.

Recent studies have begun to explore masking-based speech enhancement for spatial audio, with representations such as Ambisonics or beamformer channels~\cite{lugasi2020speech,herzog2023ambisep,9747583}. For example, \cite{lugasi2020speech} compared masking in beamformer and Ambisonics domains and showed that Ambisonics-domain masking can better preserve spatial information of both the desired signal and the residual disturbance. However, a comparison with masking applied directly to microphone channels—one of the most common signal representations—has not been systematically investigated, particularly with respect to the trade-off between speech quality, intelligibility, and spatial fidelity.

The goal of this work is to analyze masking-based speech enhancement algorithms and to investigate the trade-offs between speech enhancement performance, speech quality and intelligibility, and spatial fidelity. We compare time–frequency masking applied to three signal representations: microphone channels (TFM), beamformer outputs (TFB), and Ambisonics (TFA). In addition, we consider two cases of the desired signal: the direct sound of the target speaker and the complete reverberant target signal. Using simulated acoustic scenes with a spherical microphone array, we evaluate the methods in terms of speech quality, intelligibility, spatial cue preservation, and reverberation characteristics.

This study presents the following contributions: (i) demonstrating the need for a reverberant desired reference in the loss function to preserve reverberation; (ii) showing that masking applied to beamforming channels best preserves speech quality and intelligibility; (iii) showing that masking Ambisonics channels best preserves the spatial qualities of the residual disturbance; and (iv) showing that masking in all signal representations preserves the spatial cues of the target signal. 

\section{BACKGROUND}
\label{sec:system}
\subsection{Signal Model}
We consider a reverberant acoustic scene containing a desired speaker, interferers and stationary background noise.
The signal is assumed to be received by a spherical microphone array, represented in the frequency domain as:
\begin{equation}
\mathbf{x}(f) = \mathbf{x}^{\text{dir}}(f) + \mathbf{x}^{\text{rev}}(f) + \mathbf{n}(f)
\label{eq:mic}
\end{equation}
where $\mathbf{x}(f)$ is the $Q\times1$ vector representing the signals at the $Q$ microphones at frequency $f$. The microphones are positioned on a rigid sphere of radius $r$ at angles described in spherical coordinates as $\Phi=(\theta_q,\phi_q)$ for $q=1,...,Q$. Here, $\mathbf{x}^{\text{dir}}(f)$ denotes the signal due to the direct path of the target speaker, $\mathbf{x}^{\text{rev}}(f)$ represents its reverberant component, and $\mathbf{n}(f)$ comprises the additive noise and other interferers.

A second representation of the recorded signal can be formulated in the Ambisonics or spherical harmonics (SH) domain as described in ~\cite{lugasi2020speech}:
 \begin{equation}
\mathbf{a}_{nm}(f)=\mathbf{a}_{nm}^{\text{dir}}(f) + \mathbf{a}_{nm}^{\text{rev}}(f) + \mathbf{a}_{nm}^{\text{noise}}(f)
\label{eq:Ambisonics}
\end{equation}
where $\mathbf{a}_{nm}(f) =
[a_{00}(f), a_{1(-1)}(f),a_{10}(f), \dots, a_{NN}(f)]^T$
 is the $(N+1)^2\times1$ vector of the Ambisonics coefficients ~\cite{rafaely2015fundamentals}. $\mathbf{a}_{nm}^{\text{dir}}(f)$, $\mathbf{a}_{nm}^{\text{rev}}(f)$ and $\mathbf{a}_{nm}^{\text{noise}}(f)$ are the direct, reverberant and noise components similar to Eq. ~\eqref{eq:mic}.

The third representation is formulated in the beamformer domain:
 \begin{equation}
\mathbf{b}(f) = \mathbf{b}^{\text{dir}}(f)+\mathbf{b}^{\text{rev}}(f) + \mathbf{b}^{\text{noise}}(f)
\label{eq:pwd_stft}
\end{equation}
where $\mathbf{b}(f)$ is the $B\times1$ vector representing beamforming outputs directed toward different look directions, and $\mathbf{b}^{\text{dir}}(f)$, $\mathbf{b}^{\text{rev}}(f)$ and $\mathbf{b}^{\text{noise}}(f)$ are the direct, reverberant and noise components similar to Eqs.~\eqref{eq:mic} and ~\eqref{eq:Ambisonics}.

\subsection{Transformation Between Signal Representations}
In this work, it is assumed that the baseline signal representation is given in the Ambisonics domain as in Eq.~\eqref{eq:Ambisonics}, and the other two representations are computed from Eq.~\eqref{eq:Ambisonics} as detailed in ~\cite{rafaely2015fundamentals}. In order to transform from the SH domain to the beamformer or microphone domain, we use:
\begin{align}
\mathbf{b}(f) &= \mathbf{D}_{\text{Beam}}(f)\mathbf{a}_{nm}(f) \label{eq:decoder_Beam} \\
\mathbf{x}(f) &= \mathbf{D}_{\text{Mic}}(f)\mathbf{a}_{nm}(f)  \label{eq:decoder_mic}
\end{align}
where $\mathbf{D}_{\text{Beam}}(f)$ and $\mathbf{D}_{\text{Mic}}(f)$ are linear decoding matrices as described in ~\cite{lugasi2020speech,rafaely2015fundamentals}. To transform back to the SH domain we will use:
\begin{align}
\mathbf{a}_{nm}(f) &= \mathbf{D}^{-1}_{\text{Beam}}(f)\mathbf{b}(f) \label{eq:encoder_Beam} \\
\mathbf{a}_{nm}(f) &= \mathbf{D}^{-1}_{\text{Mic}}(f)\mathbf{x}(f)
\label{eq:encoder_mic}
\end{align}
The decoding matrices may not be invertible in general; however, invertibility is ensured in our setup (Section~\ref{sec:ANALYSIS}), though the inverse in Eq.~\eqref{eq:encoder_mic} may introduce some small error.

\subsection{Time-Frequency Masks}
After transformation from the frequency to the time domain, the measured signals in the three representations can be further transformed by applying the short-time Fourier transform (STFT), leading to $\mathbf{x}(\tau,\nu)$, $\mathbf{a}_{nm}(\tau,\nu)$ and $\mathbf{b}(\tau,\nu)$. We can estimate the desired components by multiplying each of these signals with a masking matrix:
\begin{align}
\hat{\mathbf{x}}(\tau,\nu) &= \mathbf{M}_{\text{Mic}}(\tau,\nu) \mathbf{x}(\tau,\nu) \label{eq:mask_mic} \\
\hat{\mathbf{a}}_{nm}(\tau,\nu) &= \mathbf{M}_{\text{Amb}}(\tau,\nu) \mathbf{a}_{nm}(\tau,\nu) \label{eq:mask_amb} \\
\hat{\mathbf{b}}(\tau,\nu) &= \mathbf{M}_{\text{Beam}}(\tau,\nu) \mathbf{b}(\tau,\nu) \label{eq:mask_Beam}
\end{align}
where $\mathbf{M}_{\text{Mic}}(\tau,\nu)$, $\mathbf{M}_{\text{Beam}}(\tau,\nu)$ and $\mathbf{M}_{\text{Amb}}(\tau,\nu)$ are diagonal matrices, representing ideal ratio masks (IRM) ~\cite{lugasi2020speech}. The IRM is widely used as an upper-bound oracle for mask-based methods, including DNN-based approaches~\cite{gaultier2024recovering}, since it requires knowledge of the true signal and noise components and is thus not realizable in practice. Here it serves as a reference for the theoretical performance limits of each representation. The desired signal is defined as either the direct or the reverberant target. At the end of this process, the signals are transformed back to the time domain.

\subsection{Binaural Representation}
 The enhanced signals can be decoded to binaural signals after transformation to the frequency domain, employing the Head-Related Transfer Function (HRTF):
 \begin{equation}
b^{l/r}(f) =  \tilde{\mathbf{h}}^{l/r}_{nm}(f)^T \mathbf{a}_{nm}(f)
\label{eq:Binaural}
\end{equation}
where $\tilde{\mathbf{h}}^{l/r}_{nm}(f)$ is the modified SH-domain representation of the HRTF, obtained from $\mathbf{h}^{l/r}_{nm}(f)$, where $\tilde{h}^{l/r}_{nm}(f)= (-1)^mh^{l/r}_{n(-m)}(f)$~\cite{gayer2025ambisonics}. For the beamformer and microphone domain signal, a transformation back to the SH domain is employed using Eqs.~\eqref{eq:encoder_Beam} and ~\eqref{eq:encoder_mic} before computing the binaural signal.

\section{OBJECTIVE MEASURES OF PERFORMANCE}
\label{sec:MEASURES}

Performance is evaluated using objective measures that quantify signal reconstruction accuracy, spatial cue preservation, reverberation characteristics, and speech intelligibility and quality.

Overall signal preservation is measured using the scale-invariant signal-to-distortion ratio (SI-SDR)~\cite{le2019sdr}, averaged across the binaural channels. The reference signal is either the direct target signal or the reverberant target signal depending on the experiment.

Preservation of spatial cues is evaluated using interaural time difference (ITD) and interaural level difference (ILD)~\cite{blauert2013technology,hauth2020modeling}. Performance is quantified as the absolute error between the processed and reference binaural signals, denoted $\Delta \text{ITD}$ and $\Delta \text{ILD}$ respectively.

Reverberation characteristics are assessed using the clarity measure $\text{C}_{50}$~\cite{christensen2013iso}, which represents the ratio between signal energy arriving early (0 - 50 ms delay) and late. We denote the absolute error between the processed and reference values averaged across ears as $\Delta{\text{C}_{50}}$.

Speech intelligibility is evaluated using deterministic binaural STOI (DBSTOI)~\cite{andersen2016method} which takes into account binaural spatial release from masking, with the direct target speech as the reference. Speech quality is assessed using PESQ~\cite{rix2001perceptual}, computed independently for the two ears and averaged.

\section{Simulation Study}
\label{sec:ANALYSIS}
This section evaluates the performance of the three masking domains through controlled acoustic simulations. We examine how different signal representations and reference definitions influence speech quality, spatial fidelity, and reverberation preservation across various noise and interference conditions.

\subsection{Setup}
We conducted Monte-Carlo simulations consisting of 100 realizations of the acoustic scene under various conditions of speech signals, room acoustics, SNR, DRR, and source geometry, as summarized in Table~\ref{tab:simulation_data}. In each realization, a rectangular room was simulated and a spherical microphone array was placed in the room maintaining a distance of at least $1~\text{m}$ from the walls. The target speaker was placed at a horizontal angle of $0^\circ$ relative to the array, and at varying distances to control the Direct-to-Reverberant Ratio (DRR). An interfering speaker was positioned at the same distance as the target speaker and separated by different azimuth angles. A stationary noise source producing white noise was placed in one corner of the room to simulate a highly reverberant noise field. The total noise is defined to be the sum of the interfering speaker and the stationary noise. 
The room impulse response is calculated using the image method ~\cite{allen1979image} and used to compute the sound field around the array in the spherical harmonics domain, including terms up to order $N = 35$ as in ~\cite{rafaely2015fundamentals}. Anechoic speech signals were used as source signals and convolved with the simulated RIRs to obtain $\mathbf{a}_{nm}(f)$. For each realization, the room, the geometry of the source-array and the speech signal were randomly selected as specified in Table~\ref{tab:simulation_data}. The resulting acoustic configuration was then evaluated across all SNRs, DRRs and desired signals.

\subsection{Methodology}
We consider two definitions for the desired signal $\mathbf{a}^d_{nm}(f)$ and the corresponding undesired $\mathbf{a}^u_{nm}(f)$: (i) Direct signal, with $\mathbf{a}^d_{nm}(f) = \mathbf{a}_{nm}^{\text{dir}}(f)$ and $\mathbf{a}^u_{nm}(f) = \mathbf{a}^{\text{rev}}_{nm}(f) + \mathbf{a}^{\text{noise}}_{nm}(f)$; 
(ii) Reverberant signal, with $\mathbf{a}^d_{nm}(f) = \mathbf{a}_{nm}^{\text{dir}}(f)+\mathbf{a}^{\text{rev}}_{nm}(f)$ and $\mathbf{a}^u_{nm}(f) = \mathbf{a}^{\text{noise}}_{nm}(f)$.
Ambisonics signals were truncated to order $N=3$ (16 channels), representing a practical system such as the Eigenmike32~\cite{acoustics2013em32}. Beamformer signals were generated from this representation using Eq.~\eqref{eq:decoder_Beam}, with maximum-directivity beamformers implemented as in~\cite{lugasi2020speech}. Microphone signals were instead generated from Ambisonics of order $N=10$ using Eq.~\eqref{eq:decoder_mic}, assuming a rigid spherical array with radius $r = 0.042$~m. In both cases, $Q=B=24$ sampling directions were used, distributed according to a spherical $t$-design for near-uniform spatial sampling~\cite{rafaely2015fundamentals}. Time–frequency real-valued ideal ratio masks were then computed using the known desired and undesired components as described in ~\cite{lugasi2020speech}. The masks were then applied to the corresponding signal as in Eqs.~\eqref{eq:mask_mic}, ~\eqref{eq:mask_amb} and ~\eqref{eq:mask_Beam}. Binaural signals were
finally computed by using Eq.~\eqref{eq:Binaural} with the Cologne HRTF compilation of the Neumann KU-100 ~\cite{bernschutz2013spherical}. 

\begin{table}[t]
\centering
\caption{Details of the Monte-Carlo simulation}
\label{tab:simulation_data}
\begin{tabular}{>{\raggedright\arraybackslash}p{2.0cm} >{\raggedright\arraybackslash}p{5.8cm}}
\toprule
\textbf{Variable} & \textbf{Description} \\
\midrule
Speech & Ten WSJ utterances~\cite{paul1992design}, $f_s=16$ kHz \\
\midrule
Room & Three rectangular rooms:
\newline Room 1: $10 \times 6 \times 3.2\,\text{m}$, $T_{60}=0.44$
\newline Room 2: $13 \times 8 \times 3.5\,\text{m}$, $T_{60}=0.56$
\newline Room 3: $20 \times 10 \times 4\,\text{m}$, $T_{60}=0.75$ \\
\midrule
SNR & Target–interferer: $\{-20,-10,0,10\}\,$dB; target–noise: $20\,$dB \\
\midrule
DRR & $\{-5, 10\}\,$dB \\
\midrule
Positions & Target--interferer angle: $\{-100,60\}^\circ$ \\
\midrule
Desired Signal & Direct signal or fully reverberant signal of the target speaker \\
\bottomrule
\end{tabular}
\end{table}

\subsection{Results}
The following results examine how each masking domain performs across four aspects: reverberation preservation, speech quality and intelligibility, and spatial attribute preservation of both the desired and interfering signals.

\subsubsection{Preservation of Late Reverberation}
Table ~\ref{tab:C50_table} shows $\Delta \text{C}_{50}$ when the desired signal is either the direct signal or the complete reverberant signal of the target speaker, both compared to the reverberant signal of the target speaker. For the direct desired signal, all masking approaches exhibit $\Delta \text{C}_{50}$ larger than the just noticeable difference (JND) which is approximately 1 dB ~\cite{christensen2013iso}, with TFB exhibiting the highest difference. This is expected, since using the direct path as the desired signal inherently suppresses late reverberation, increasing $\text{C}_{50}$ regardless of masking domain. In contrast, for the reverberant desired signal, all masks preserve the reverberation level relatively well with the $\Delta \text{C}_{50}$ of the TFA and TFM methods lower than the JND, while the TFB method is least accurate. It is therefore quite clear that when aiming to preserve spatial environmental attributes such as reverberation level, using the direct signal as the desired signal is not useful. Therefore, in the following analysis we will focus on the reverberant target as the desired signal. 

\begin{table}[h]
    \centering
    \caption{$\Delta{\text{C}_{50}}$ in dB (lower is better) for the three masking methods, using either the direct signal or the reverberant signal as the desired signal in the masking process. The average $\text{C}_{50}$ under all channels was $3.43\,$dB with the low DRR, and $12.92\,$dB for the high DRR.}
    \label{tab:C50_table}
    \normalsize
    \begin{tabular}{lcccc}
        \toprule
        \textbf{Method} 
        & \multicolumn{2}{c}{\textbf{DRR = -5 dB}} 
        & \multicolumn{2}{c}{\textbf{DRR = 10 dB}} \\
        \cmidrule(lr){2-3} \cmidrule(lr){4-5}
        & \textbf{\textbf{Direct}} 
        & \textbf{\textbf{Reverberant}} 
        & \textbf{\textbf{Direct}} 
        & \textbf{\textbf{Reverberant}} \\
        \midrule
        
        TFM  & 3.00 & 0.41 & 2.91 & 0.92 \\
        TFB  & 9.00 & 0.58 & 7.95 & 1.63 \\
        TFA & 3.72 & 0.40 & 3.20  & 0.82 \\
        
        \bottomrule
    \end{tabular}
\end{table}

\subsubsection{Preservation of Speech Quality and Intelligibility of the Desired Signal}
Table ~\ref{tab:reverberant_target_average_table} shows the values of the measures related to the desired signal averaged over the entire Monte-Carlo dataset when the desired signal is the reverberant target. It shows that on average, TFB achieves the highest SI-SDR and PESQ scores indicating superior target signal reconstruction and speech quality, as well as the highest DBSTOI score, reflecting better intelligibility. On the other hand, TFM yields the lowest SI-SDR, PESQ and  DBSTOI scores indicating the poorest estimation performance. The superior performance of TFB and TFA over TFM can be explained by the spatial pre-processing providing some spatial separation through the SH patterns in TFA and the maximum-directivity patterns in TFB.

\begin{table}[t]
\centering
\caption{Performance measures with respect to the desired signal, for the three masking methods using the reverberant signal as the desired signal in the
masking process, averaged over all scenarios}
\label{tab:reverberant_target_average_table}

\resizebox{\columnwidth}{!}{
\large
\begin{tabular}{lcccccc}
\toprule
\textbf{Method} 
& \shortstack{\textbf{SI-SDR} \\ (dB) $\uparrow$} 
& \shortstack{\textbf{$\Delta$ITD} \\ ($\mu$s) $\downarrow$} 
& \shortstack{\textbf{$\Delta$ILD} \\ (dB) $\downarrow$} 
& \shortstack{\textbf{$\Delta$C50} \\ (dB) $\downarrow$} 
& \shortstack{\textbf{DBSTOI} \\ (-) $\uparrow$} 
& \shortstack{\textbf{PESQ} \\ (-) $\uparrow$} \\
\midrule
Noisy  
& -6.37 & - & - & - & 0.478 & 1.26 \\
\midrule
TFM  
& 5.73 & 0 & 0.49 & 0.66 & 0.700 & 2.49 \\
\midrule
TFA 
& 9.36 & 0 & 0.07 & \textbf{0.61} & 0.745 & 2.66 \\
\midrule
TFB  
& \textbf{12.93} & 0 & 0.40 & 1.10 & \textbf{0.802} & \textbf{3.09} \\
\bottomrule
\end{tabular}
}
\end{table}

\subsubsection{Preservation of the Desired Signal Spatial Attributes}
Table ~\ref{tab:reverberant_target_average_table} also shows that, although small numerical differences exist between methods, all $\Delta \text{ITD}$ and $\Delta \text{ILD}$ for the target speaker remain well below perceptual thresholds ~\cite{spencer2016relating,blauert2013technology}, of about 100 $\mu$s for ITD and 1 dB for ILD, indicating that when the desired signal is the reverberant target, all methods preserve the binaural spatial cues of the target speaker within perceptual limits. As for $\Delta \text{C}_{50}$, these values are consistent with Table \ref{tab:C50_table}.

\subsubsection{Preservation of the Residual Interfering Signal Spatial Attributes}
Table ~\ref{tab:reverberant_interfering_average_table} shows the values of the measures related to the interfering signal averaged over the entire Monte-Carlo dataset when the desired signal is the reverberant signal. All the methods have $\Delta \text{ITD}$ and $\Delta \text{ILD}$ above the JND threshold but the difference is lowest for TFA, especially the $\Delta \text{ITD}$. TFM's reduced performance may stem from high channel correlation and interference contribution to all microphones; TFB is similarly affected by beampattern sidelobes, whereas TFA's orthogonal beam patterns may better mitigate this leakage. These results indicate that TFA better preserves the spatial attributes of the interfering signal compared to the other methods and may enable SRM and better scene awareness.

\begin{table}[h]
\centering
\caption{Performance measures, with respect to the interference signal for the three masking methods using the reverberant signal as the desired signal in the
masking process, averaged over all scenarios}
\label{tab:reverberant_interfering_average_table}
\begin{tabular}{lcc}
\toprule
\textbf{Method} 
& \textbf{$\Delta$ITD ($\mu$s) $\downarrow$} 
& \textbf{$\Delta$ILD (dB) $\downarrow$} \\
\midrule
Noisy  
& - & - \\
\midrule
TFM  
& 463 & 6.67 \\
\midrule
TFA 
& \textbf{43} & \textbf{3.31} \\
\midrule
TFB  
& 186 & 7.72 \\
\bottomrule
\end{tabular}
\end{table}

\section{CONCLUSIONS}
This paper compares time-frequency masking-based speech enhancement across three representations. The analysis reveals a distinct trade-off: while all methods preserve the target speaker's spatial cues (ITD/ILD) within perceptual limits, Ambisonics channels best preserve the spatial attributes of the residual interference and the acoustic scene ($\text{C}_{50}$), whereas beamformer channels provide superior noise suppression and speech quality, at the expense of spatial fidelity. Notably, microphone channels proved least effective for masking, due to the lack of spatial separation in the raw channels. Analysis also showed that defining the reverberant target as the desired signal significantly improved the preservation of $\text{C}_{50}$. 
Future work is proposed to extend this analysis to practical systems, where masks are estimated using deep neural networks, and to incorporate listening tests to assess spatial quality and speech intelligibility. Additionally, the analysis should be extended to first-order Ambisonics and to non-ideal array configurations toward more practical deployment scenarios.

\label{sec:Conclusions}

\clearpage 

\begingroup
\small
\balance
\bibliographystyle{IEEEbib}
\bibliography{refs}
\endgroup

\end{document}